\documentclass[11pt]{article}
\usepackage{amsmath,multirow}

\usepackage{ifpdf}
\ifpdf
    \usepackage[pdftex]{graphicx}   
    \usepackage[pdftex,     
            plainpages=false,   
            breaklinks=true,    
            colorlinks=true,
            pdftitle=My Document
            pdfauthor=My Good Self
           ]{hyperref}
    \usepackage{thumbpdf}
\else
    \usepackage{graphicx}       
    \usepackage{hyperref}       
\fi

\title{MCA: Boolean Networks Control Algorithm}
\author{Mohammad Moradi\thanks{University of Tehran, Tehran, Iran.} \and Sama Goliaei\footnotemark[1] \\
}

\begin{document}
\maketitle
\begin{abstract}
   Control problem in a biological system is the problem of finding an interventional policy for changing the state of the biological system from an undesirable state, e.g.  disease, into a desirable healthy state. Boolean networks are utilized as mathematical model for gene regulatory networks. This paper provides an algorithm to solve the control problem in Boolean networks. The proposed algorithm is implemented and applied on two biological systems: T-cell receptor network and Drosophila melanogaster network. Results show that the proposed algorithm works faster in solving the control problem over these networks, while having similar accuracy, in comparison to previous methods.


  \textbf{Highlights}:
  \begin{itemize}
    \item Using dynamic programming method to reduce time of the Boolean network control problem
    \item Concentration on branching nodes, with accessibility to both states of 0 and 1 in the same time steps
    \item The less number of branching nodes with accessibility to both states  0 and 1 in the same time steps, the less time consumption
    \item Quicker detection of lack of a control sequence in most of the cases
  \end{itemize}
\end{abstract}

\section{Introduction}
A gene regulatory network (GRN) is a set of genes and relations between them~\cite{alberts2002molecular}. The purpose of GRNs mathematical modelling is to achieve a new insight toward the important cellular processes. As instances of mathematical modelling of biological processes, we can refer to cell cycle~\cite{li2004yeast,tyson2001network}, oscillations in p53-mdm2 system~\cite{batchelor2009ups,choi2012attractor,geva2006oscillations}, phage-lambda system~\cite{joh2011lyse,murrugarra2012modeling,zeng2010decision}, and T-cell large granular lymphocyte (T-LGL) leukemia network~\cite{saadatpour2011dynamical,zhang2008network}. There exist different techniques for modelling dynamics of GRNs, including Boolean networks (BNs)~\cite{zhang2008network}, Bayesian networks~\cite{zhang2008network}, dynamic Bayesian networks~\cite{zhang2008network}, linear models~\cite{shih2004prediction} and differential equations~\cite{shih2004prediction}. Among the above mentioned models, the Boolean network model has received many attentions~\cite{shih2004prediction,stuart1993origins,akutsu2000inferring,albert2000dynamics,amaral2004emergence,harris2002model,liang1998reveal,pandey2010boolean,helikar2008emergent,kim2011reduction}; that is because in addition to the tractability, Boolean networks could be reconstructed by efficient biological experiments~\cite{helikar2011boolean,bornholdt2008boolean}.

Detecting a set of perturbations which cause the desirable changes in cellular behavior has many applications such as cancer treatment and drug discovery~\cite{kitano2002computational,kitano2004cancer,choi2012attractor,erler2012network,lee2012sequential,wang2013therapeutic}. This highlights the necessity of developing a control theory for the gene regulatory networks. Using GRN control mathematical models is considered as a key method to design the experimental control policies~\cite{wang2013therapeutic}.

The control problem includes finding a sequence of interventions to be applied on the system, which changes state of the system from an undesirable state of the network to a desirable one~\cite{datta2003external,datta2004external,pal2005intervention}. The undesirable state in a gene regulatory network may express a disease such as cancer, and the desirable state can express the wellness, for example as induction of apoptosis in cancerous cells or tumours. Therefore, using the case of control and medical interventions, we can exterminate the tumour cells and achieve healthiness~\cite{pal2005intervention}.

Up to now, numerous methods have been proposed to solve the control problem in Boolean networks. Among the vast diverse proposed methods, we name the more optimised control techniques, to which a list of possible control nodes are given as input~\cite{yousefi2012optimal,yousefi2013intervention,yousefi2014comparison,yousefi2013optimal}. Bo Gao et al proposed an algebraic method to solve the control problem and used the semi-tensor product (STP) as a state transition matrix~\cite{gao2013principle}. Qiu, Yushan et al took benefit from the integer programming to solve the control problem in multiple Boolean networks, for the cancer-causing and normal cells~\cite{qiu2014control}. Christopher James Langmead and Sumit Kumar Jha proposed an algorithm based on model checking to find the control strategy in Boolean networks~\cite{langmead2009symbolic}. Yang Liu et al searches for a controlling sequence to transform from a state to a desirable one, with a difference that he avoids some special and prohibited states~\cite{liu2014controllability}.

Meanwhile, in some cases, the genetic algorithm and the greedy algorithms are used to solve the control problem in Boolean networks~\cite{poret2014silico,vera2013ocsana,kim2013discovery}. Datta et al proposed an algorithm to control the probabilistic Boolean networks (PBN) based on Markov chains and dynamic programming~\cite{datta2003external,datta2004external,pal2005intervention}. In this approach, it is supposed that states of some nodes could be controlled externally, and the goal is to find a sequence of changes to be applied as controlling policy to result in the network desirable state. Since the Boolean networks are a special mode of probabilistic Boolean networks, this method is also applicable on Boolean networks. The problem with the proposed algorithm was that it lacks the necessary efficiency, because all the states within probabilistic Boolean network (or Boolean network) were required to be taken into consideration in all time steps between the initial state and the desirable state; so a state transition matrix with exponential size was produced programmatically~\cite{datta2004external}.

According to studies by Akutsu et al, it was specified that finding the control strategy in Boolean networks is NP-hard. However, they proposed a polynomial time algorithm to find the control strategies over trees instead of general graphs, in which the dynamic programming was used to find the control sequence~\cite{akutsu2007control}. They also expanded their algorithm for the networks with low number of loops, but if the network has a high number of loops, or the given number of time step between the initial state and the desirable state is high, this algorithm would not have the desirable efficiency~\cite{akutsu2007control}. Meanwhile, in most of proposed methods, if the size of Boolean network is high, the proposed algorithm might have not the desirable efficiency.

Although an algorithm with polynomial time has been suggested for the networks with tree structure~\cite{akutsu2007control}, this case may also not be applicable, since most of biological networks lack a tree structure. Therefore, new algorithms is still needed which be efficient for general structure of networks whether with high number of loops, with high number of time step, and for networks with large sizes. In this paper, we have presented a new algorithm to solve the control problem. Also the applicability of the provided algorithm on the two biological systems T-cell receptor network~\cite{klamt2006methodology} and Drosophila melanogaster network~\cite{albert2004boolean}, is shown and was compared with other algorithms.

\section{Materials and Methods}

\subsection{Background on Boolean Network Control}

In a Boolean network, each node represents a gene, and each edge represents a regulatory effect of one gene expression on another one, which may cause increase or decrease in the gene expression~\cite{bar2004analyzing}.

A Boolean network is illustrated through a directed graph $G=(V,F)$, which includes a set of $n$ nodes $V = \{v_1, v_2,\dots, v_n\}$, and a set of Boolean functions $F = \{f_1, f_2, \dots, f_n\}$. Time is considered as discrete intervals in this model. Each node $v_i$ has a state variable $v_i(t)\in\{0, 1\}$ representing the state of node $v_i$ at time $t$, where $0$ ($1$) indicates lack of (existence of) gene expression. Also, each node $v_i$ has a Boolean function $f_i$, representing how to obtain $v_i(t+1)$ from the state of the incoming nodes to $v_i$ at time step $t$ by applying basic Boolean operations ({\bf{and}}, {\bf{or}}, {\bf{not}}).
The network state at time $t$, is defined as vector $v^t=[v_1(t), v_2(t), \dots, v_n(t)]$, which describes the state of the nodes in time step $t$.

\begin{figure}
  \includegraphics [width=5 in] {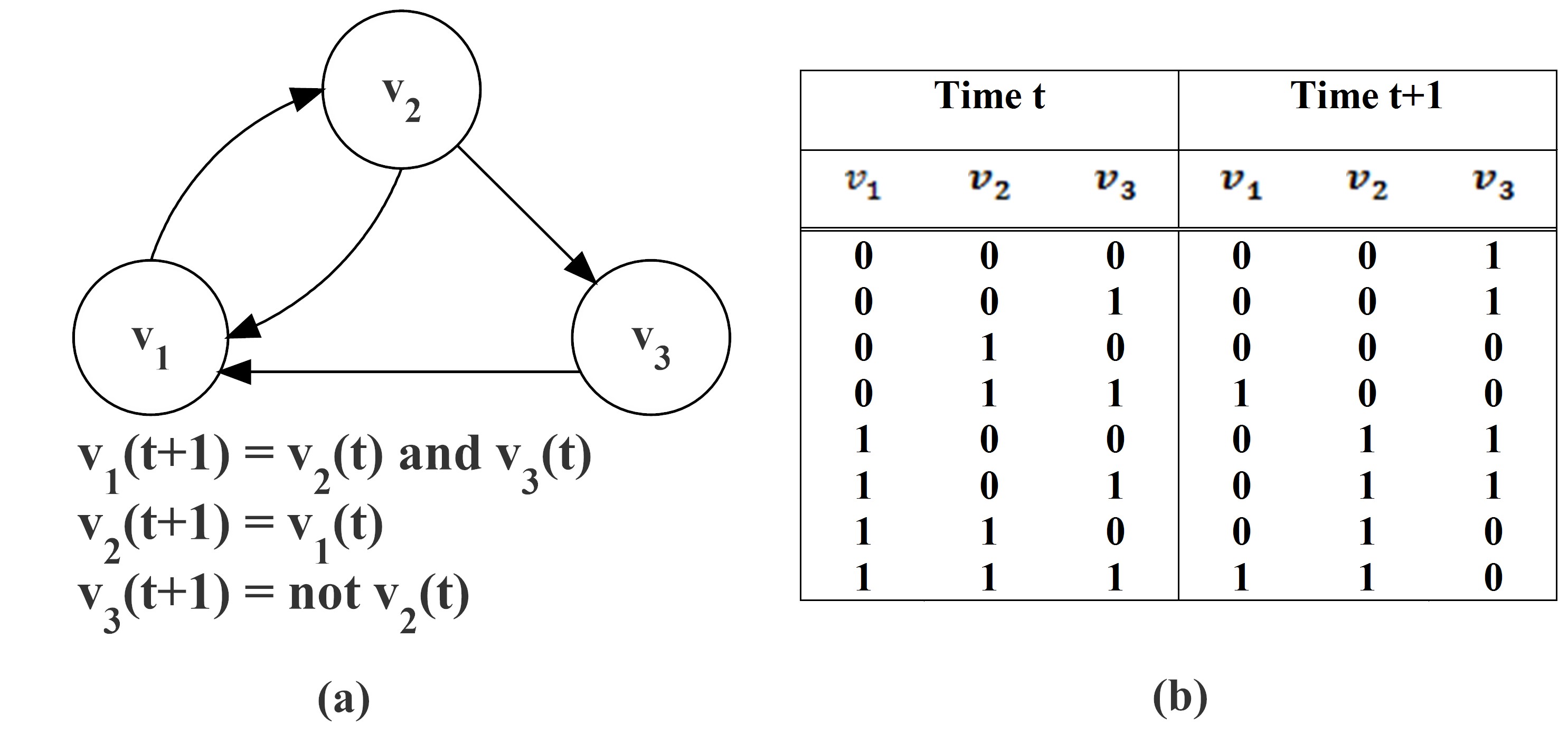}
  \caption{(a) An example of a Boolean network. (b) the state transition table of this Boolean network.}\label{fig:fig1}
\end{figure}

 An example of a Boolean networks is represented in Fig.~\ref{fig:fig1}(a). In Fig.~\ref{fig:fig1}(b), the state transition table of the mentioned Boolean network is represented. This table represents the next state of the network according to the current state. For example, if the network state in time step $t$ is $[0, 1, 1]$, then the network state in time step $t+1$ is $[1, 0, 0]$.

In the control problem of Boolean networks, a Boolean network $G=(V,F)$, initial state and a desirable network state $v^{\tau}$ is given. The set of nodes $V$ is called internal nodes. A set of control nodes $\{u_1,\dots,u_m\}$ are added to the network, known also as external nodes, which are used to influence internal nodes to attain the desirable state. The external nodes have no incoming edges, and their values are specified externally. The problem is to find a sequence of state values $u^0, \dots, u^\tau$ for external nodes, which results the network to be in desirable state $v^{\tau}$ in time step $\tau$. If there exists no such control sequence, this fact should be announced as the output. In gene regulatory networks, the desirable state of the network represents a healthy state of the system, and external nodes represent potential medicines affecting network behaviour. Thus, finding control strategies has applications in various medical areas, including medical protocol design for example in cancer treatment~\cite{kitano2002computational,kitano2004cancer}.

\begin{figure}
  \includegraphics[width=5 in]{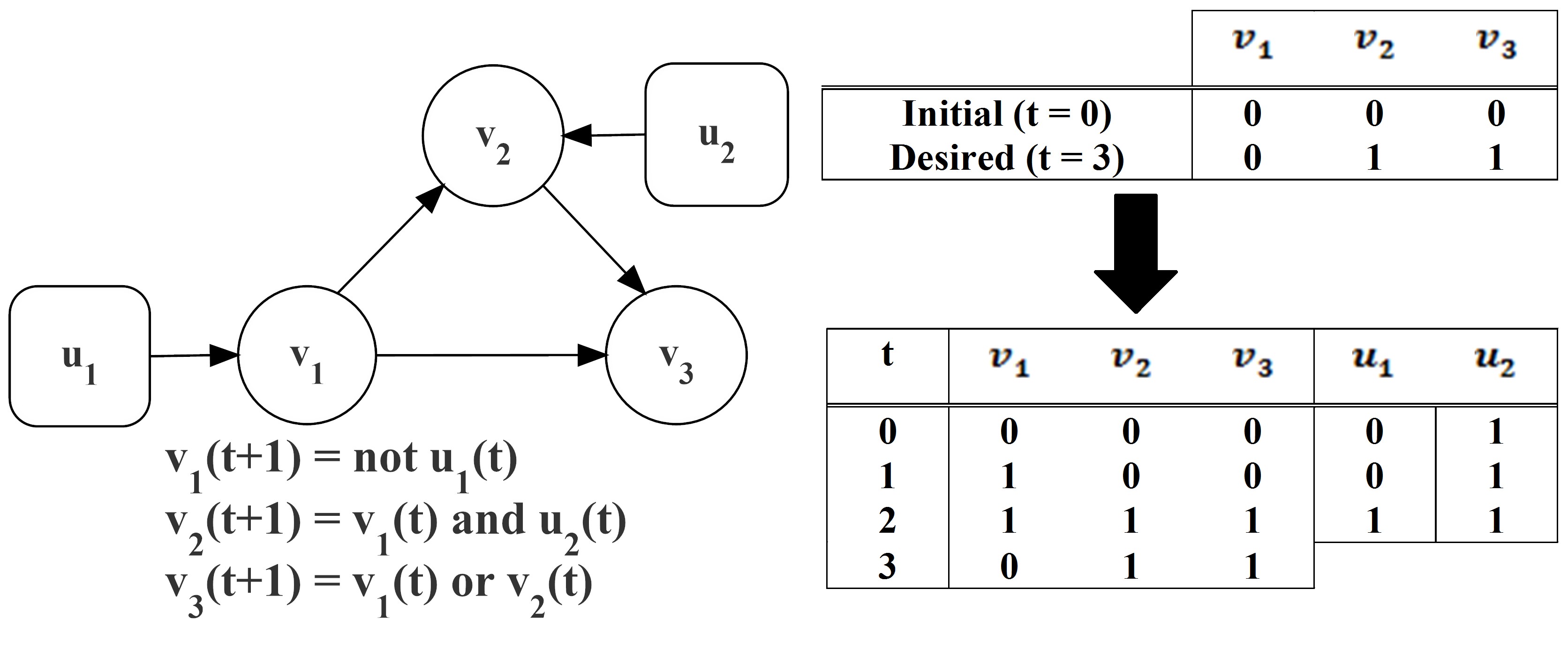}
  \caption{An example of the control problem on a Boolean network.}\label{fig:fig2}
\end{figure}

An example of a control problem on a Boolean network is represented in Fig.~\ref{fig:fig2}. In this example, $\{v_1, v_2, v_3\}$ is the set of internal nodes and $\{ u_1, u_2\}$ is the set of external nodes. The initial state of the network is $v^0=[0, 0, 0]$, and the desirable state is $v^3=[0,1,1]$. Thus, we are looking to find a control sequence $u^0,\dots, u^2$ in such a way that the network would be in state $v^3$ at time step $t=3$. A possible solution is $u^0=[0,1]$, $u^1=[0,1]$, $u^2=[1,1]$.

\subsection{Proposed Algorithm}

In our proposed algorithm, we compute some intermediate variables to be used in computation of the final result.
As intermediate variables, for each node, at each time step, we find that is it possible for this node to be in state 1, and is it possible to be in state 0? For this computation we design following steps.

For network node $v_i$, Boolean variable $b\in\{0,1\}$, and time step $t$, we define variable $\Upsilon^{v_i}_b(t)$, which is \textit{true} if and only if it is possible to assign values to external nodes in such a way that it cause $v_i(t)$ to get value $b$, and is \textit{false} otherwise. In the other words, $\Upsilon^{v_i}_1(t)$ and $\Upsilon^{v_i}_0(t)$ represent if it is possible for node $v_i$ in time step $t$ to have state value $1$ and $0$, respectively.

$\Upsilon^{v_i}_1(t+1)=true$ if, and only if, there exists $[b_{i_1}, b_{i_2}, \dots, b_{i_k}]$ such that $f_i(b_{i_1}, b_{i_2}, \dots, b_{i_k}) = 1$ and $\Upsilon^{v_{i_j}}_{b_{i_j}}(t)=true$ for all $j=0, \dots, k$, and of course it is possible to have $\Upsilon^{v_{i_j}}_{b_{i_j}}(t)=true$ for all $j=0, \dots, k$ for one setting of external nodes.
$\Upsilon^{v_i}_0(t+1)=true$, is computed the same way as the above process.

\begin{figure}
  \includegraphics[width=1.5 in]{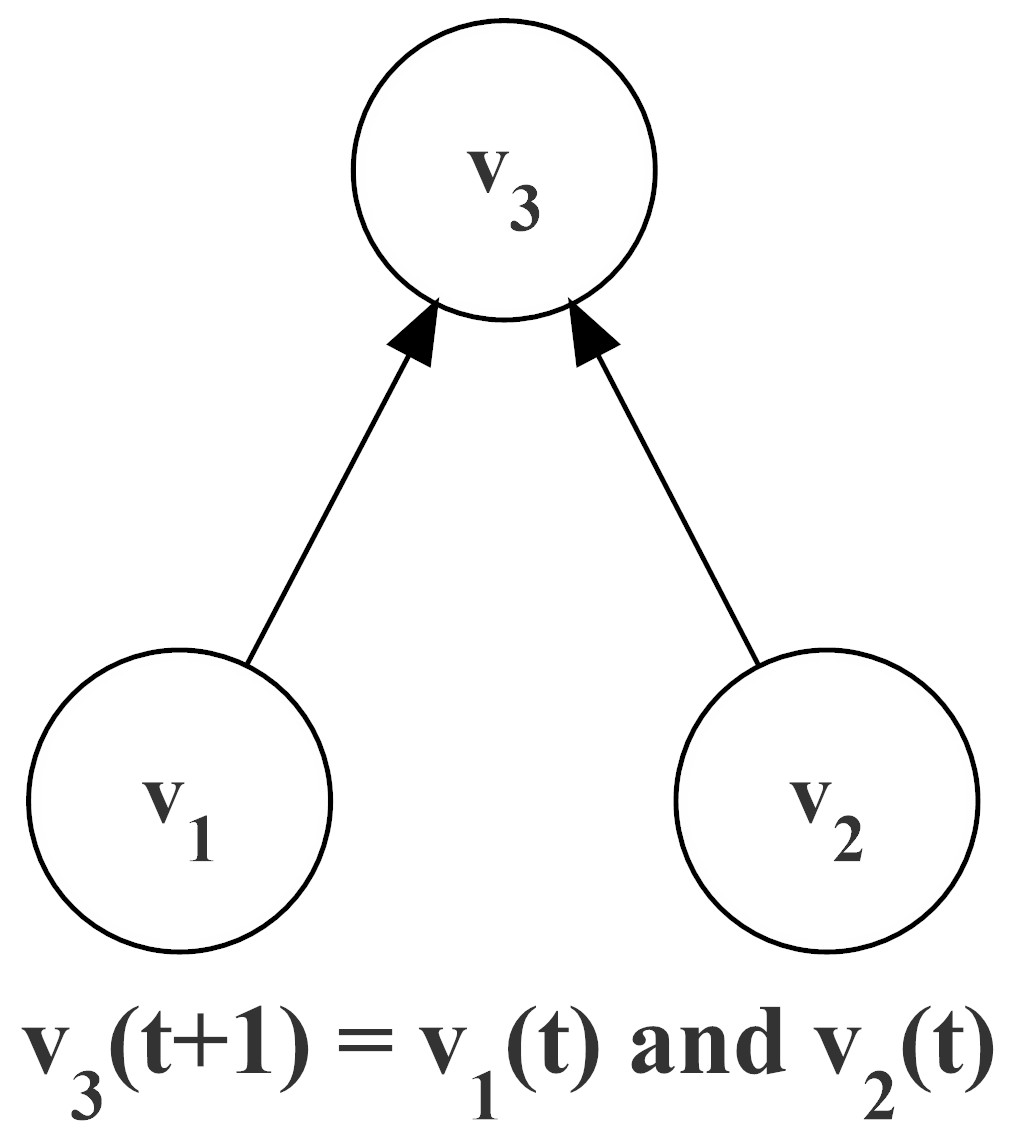}
  \centering
  \caption{Example of dynamic programming for the calculation of $\Upsilon^{v_3}_b(t+1)$. In this form $ \Upsilon^{v_3}_1(t+1)=true$, if and only if $ \Upsilon^{v_1}_1(t)=true$ and $ \Upsilon^{v_2}_1(t)=true$. Also $ \Upsilon^{v_3}_0(t+1)=true$ if and only if $ \Upsilon^{v_1}_0(t)=true$ or $ \Upsilon^{v_2}_0(t)=true$.}\label{fig:fig3}
\end{figure}

For example, as it can be seen in Fig.~\ref{fig:fig3}, the next state of node $v_3$, would be 1, if and only if the current states of the nodes $v_1$ and $v_2$ are 1. The next state of $v_3$ node would be 0, if and only if the current states of the nodes $v_1$ and $v_2$ are 0.

It must be noted that each constant node (an internal node without entering edges), or external node, could be considered as leaf nodes. For constant nodes, each of $\Upsilon^{v_i}_1(t)=true$ and $\Upsilon^{v_i}_0(t)=false$ or $\Upsilon^{v_i}_1(t)=false$ and $\Upsilon^{v_i}_0(t)=true$ are true for all time steps. Also, for each external node $v_i$, $\Upsilon^{v_i}_1(t)=true$ and $\Upsilon^{v_i}_0(t)=true$ are true for all the time steps.

To check the existence of a control sequence, $\Upsilon^{v_i}_{v^{\tau}[i]}(\tau)=true$ shall be assessed for each node. Also, to specify and output the desired control sequence, the regression technique may be used~\cite{akutsu2007control}.
In order to compute $\Upsilon$ values, we partition network into strongly connected components.

\subsubsection{Strongly Connected Components}

A \textbf{strongly connected component} (SCC) in a network, is a maximal subset of network nodes, where every node is reachable from every other in that component.
We partition the network nodes into strongly connected components, using SCC algorithm~\cite{cormen2009introductiontoalgorithm}.
A \textbf{topological order} on strongly connected components of a network, is an order on its components, which for every directed edge $xy$ from component $x$ to component $y$, $x$ comes before $y$ in the ordering.

We find a topological order on the components using topological sort algorithm~\cite{cormen2009introductiontoalgorithm}.
Partition components into three categories \emph{non-branching single node components}, \emph{branching single node components}, \emph{multi-node components}.
A branching node is the node with at least two outgoing edges and a non-branching node is the node with at most one outgoing edge.

The idea is to divide the graph into strongly connected component, and use topological sort to find a topological order on component. Then, we deal with strongly connected component according to their topological order, and treat each strongly connected component according to its type.
We assign to each strongly connected component one of the following three types.

\subsubsection{Non-Branching Single Node Components}
In this case, current component consists only of one node $v_i$, and $v_i$ has at most one outgoing edge.
We simply find $\Upsilon^{v_i}_0(t)$ and $\Upsilon^{v_i}_1(t)$ for $0\leq t\leq \tau$ from state values of incoming nodes to $v_i$ in last time step.

\begin{figure}
  \includegraphics[width=3in]{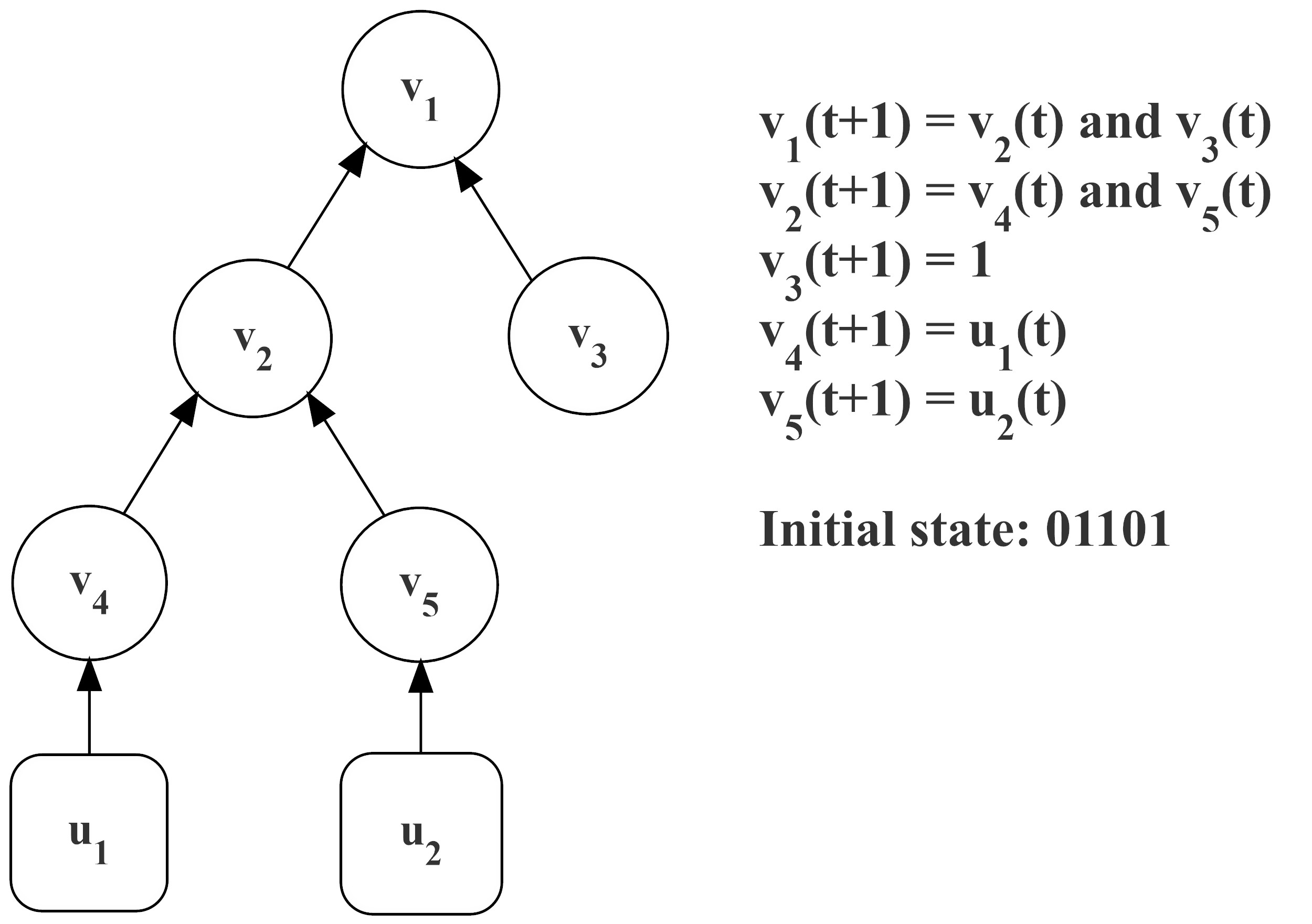}
  \centering
  \caption{An example of dynamic programming for \emph{non-branching single node component} $v_2$. Nodes $u_1$ and $u_2$ are external nodes.}\label{fig:fig4}
\end{figure}

For example, node $v_2$ in Fig.~\ref{fig:fig4} is a \emph{non-branching single node component} for which $\Upsilon^{v_2}_0(t)$ and $\Upsilon^{v_2}_1(t)$ for $0\leq t\leq 2$ is calculated based on state values of their incoming nodes $v_4$ and $v_5$ in previous time step, respectively. Values of variables $\Upsilon$ for this network is shown in Fig.~\ref{fig:UpsilonComputationOfFig4}. Note that, interestingly, for node $v_2$ in time step $t=2$ it is possible to be in state 1 and state 0.

\begin{figure}
\[
\begin{array}{l|ll}
& b=0 & b=1 \\ \hline \hline
  \multirow{7}{*}{t = 0: }
&\Upsilon^{u_1}_0(0)=true, & \Upsilon^{u_1}_1(0)=true \\
&\Upsilon^{u_2}_0(0)=true, & \Upsilon^{u_2}_1(0)=true \\
&\Upsilon^{v_1}_0(0)=true, & \Upsilon^{v_1}_1(0)=false \\
&\Upsilon^{v_2}_0(0)=false,& \Upsilon^{v_2}_1(0)=true \\
&\Upsilon^{v_3}_0(0)=false,& \Upsilon^{v_3}_1(0)=true \\
&\Upsilon^{v_4}_0(0)=true,& \Upsilon^{v_4}_1(0)=false \\
&\Upsilon^{v_5}_0(0)=false,& \Upsilon^{v_5}_1(0)=true \\ \hline
  \multirow{7}{*}{t = 1: }
&\Upsilon^{u_1}_0(1)=true,& \Upsilon^{u_1}_1(1)=true \\
&\Upsilon^{u_2}_0(1)=true,& \Upsilon^{u_2}_1(1)=true \\
&\Upsilon^{v_1}_0(1)=false,& \Upsilon^{v_1}_1(1)=true \\
&\Upsilon^{v_2}_0(1)=true,& \Upsilon^{v_2}_1(1)=false \\
&\Upsilon^{v_3}_0(1)=false,& \Upsilon^{v_3}_1(1)=true \\
&\Upsilon^{v_4}_0(1)=true,& \Upsilon^{v_4}_1(1)=true \\
&\Upsilon^{v_5}_0(1)=true,& \Upsilon^{v_5}_1(1)=true \\ \hline
  \multirow{7}{*}{t = 2: }
&\Upsilon^{u_1}_0(2)=true,& \Upsilon^{u_1}_1(2)=true \\
&\Upsilon^{u_2}_0(2)=true,& \Upsilon^{u_2}_1(2)=true \\
&\Upsilon^{v_1}_0(2)=true,& \Upsilon^{v_1}_1(2)=false \\
&\Upsilon^{v_2}_0(2)=true,& \Upsilon^{v_2}_1(2)=true \\
&\Upsilon^{v_3}_0(2)=false,& \Upsilon^{v_3}_1(2)=true \\
&\Upsilon^{v_4}_0(2)=true,& \Upsilon^{v_4}_1(2)=true \\
&\Upsilon^{v_5}_0(2)=true,& \Upsilon^{v_5}_1(2)=true \\
\end{array}
\]
  \caption{Example for computing values of variables $\Upsilon$ for \emph{non-branching single node component} of network of Fig.~\ref{fig:fig4}.}\label{fig:UpsilonComputationOfFig4}
\end{figure}

Therefore taking benefit from dynamic programming, fill $\Upsilon$ till desirable time step $\tau$. (Since the nodes are met based on a topological sort, and all the states of the incoming nodes of the related node till the time step $\tau$ are calculated before, this would be possible). In time step $\tau$, check if the related node has attained the desirable state or not. If the answer was negative, announce that there is no control sequence.

\subsubsection{Branching Single Node Components}
In this case, current component consists only of one node $v_i$, and $v_i$ has at least two outgoing edges. Same as the previous case, we can simply find values $\Upsilon^{v_i}_0(t)$ and $\Upsilon^{v_i}_1(t)$.
The difference is that since $v_i$ has more than two outgoing edges, there are at least two other nodes in the network which their state value depends on the state value of $v_i$ in the previous time. Thus, in some networks, the state value of $v_i$ which is required by its outgoing neighbours may be inconsistent.

\begin{figure}
  \includegraphics[width=2in]{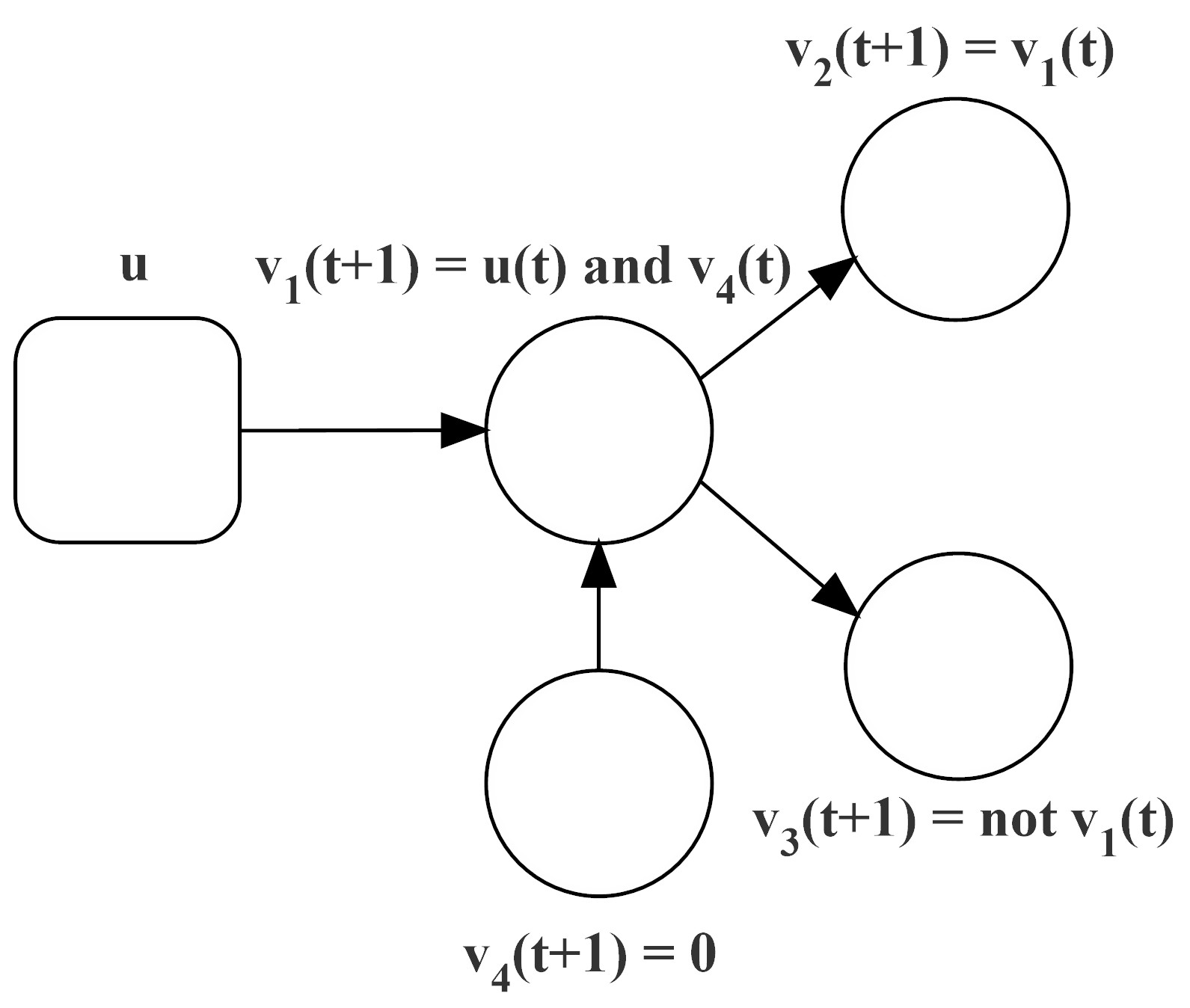}
  \centering
  \caption{An example of a branching node that is not able to attain both states 0 and 1 in a time step.}\label{fig:fig5}
\end{figure}

We first check for the current component in a time step whether is it possible to be in both states 0 and 1, or not.
For example in Fig.~\ref{fig:fig5}, the branching node $v_1$ regardless of applied control sequence is only able to attain one value (value 0) in every time step. That is because ``$v_1(t+1)=u(t) \text{\bf{and}} v_4(t)$'' and $v_4(t)$ is constantly being at state 0. We treat this node as \emph{non-branching single node components}.

\begin{figure}
  \includegraphics[width=2in]{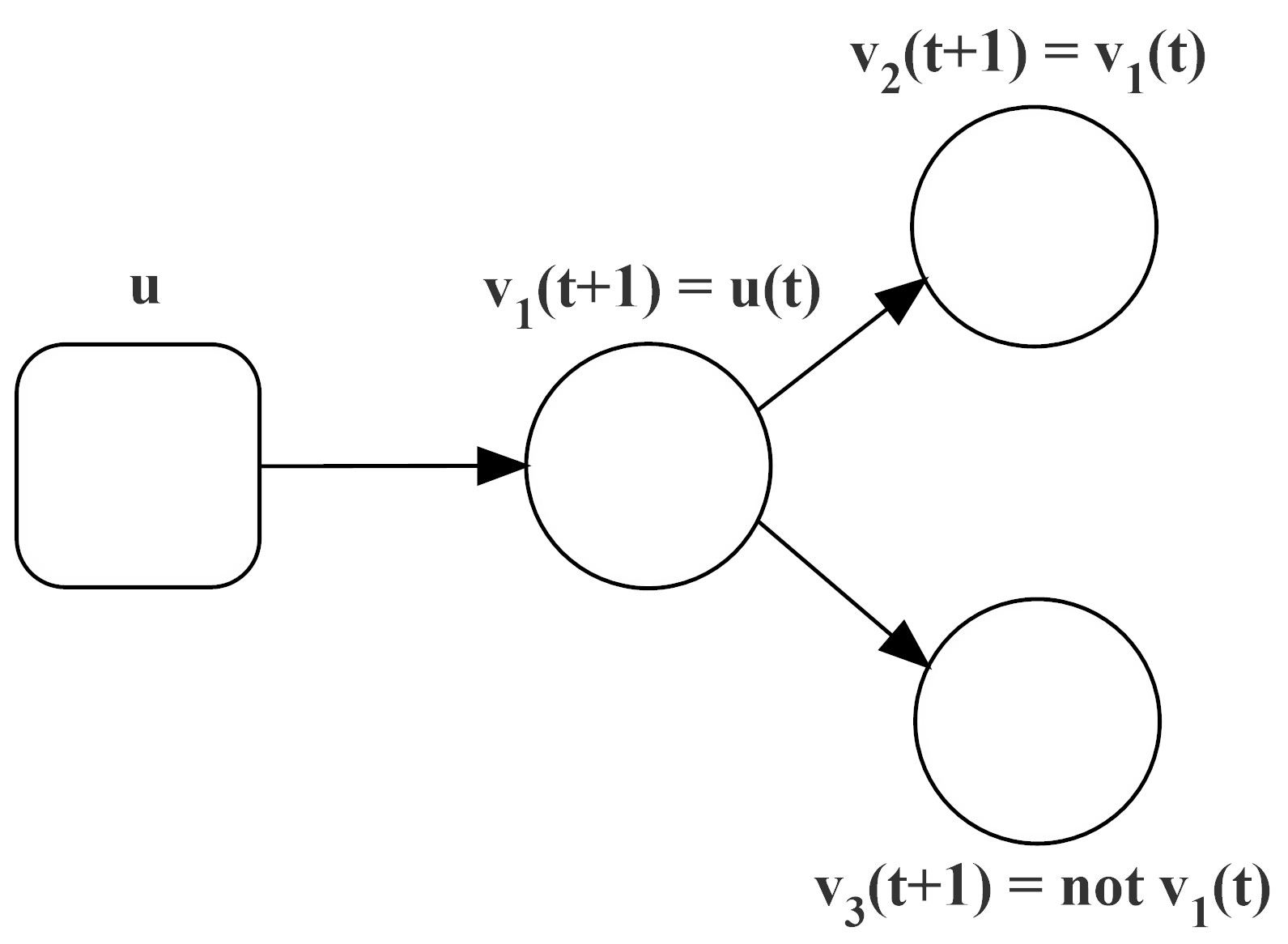}
  \centering
  \caption{An example of a branching node that is possible to attain both states 0 and 1 in a time step.}\label{fig:fig6}
\end{figure}

Some branching nodes are possible to attain both states 0 and 1 in a time step, for example, consider the network in Fig.~\ref{fig:fig6}. In this figure $u$ is an external node and $v_1$, $v_2$ and $v_3$ are internal nodes.
Suppose that initial state is $000$ and we are supposed to reach to state $111$ in time step $t = 2$. If we treat this node same as \emph{non-branching single node components}, we would face a problem. This algorithm announces that in time step $t = 2$, we can attain state $111$, but there is no a such correct control sequence. Accordingly, we fill $\Upsilon$ for this network as it is shown in Fig.~\ref{fig:UpsilonComputationOfFig6}.

\begin{figure}
\[
\begin{array}{l|ll}
& b=0 & b=1 \\ \hline \hline
  \multirow{4}{*}{t = 0: }
&\Upsilon^{u}_0(0)=true,&  \Upsilon^{u}_1(0)=true \\
&\Upsilon^{v_1}_0(0)=true,& \Upsilon^{v_1}_1(0)=false \\
&\Upsilon^{v_2}_0(0)=true,& \Upsilon^{v_2}_1(0)=false \\
&\Upsilon^{v_3}_0(0)=true,& \Upsilon^{v_3}_1(0)=false \\ \hline
  \multirow{4}{*}{t = 1: }
&\Upsilon^{u}_0(0)=true,&  \Upsilon^{u}_1(0)=true \\
&\Upsilon^{v_1}_0(1)=true,&  \Upsilon^{v_1}_1(1)=true \\
&\Upsilon^{v_2}_0(1)=true,&  \Upsilon^{v_2}_1(1)=false \\
&\Upsilon^{v_3}_0(1)=false,& \Upsilon^{v_3}_1(1)=true \\ \hline
  \multirow{4}{*}{t = 2: }
&\Upsilon^{u}_0(2)=true,& \Upsilon^{u}_1(0)=true \\
&\Upsilon^{v_1}_0(2)=true,& \Upsilon^{v_1}_1(2)=true \\
&\Upsilon^{v_2}_0(2)=true,& \Upsilon^{v_2}_1(2)=true \\
&\Upsilon^{v_3}_0(2)=true,& \Upsilon^{v_3}_1(2)=true \\
\end{array}
\]
  \caption{Example of computing values of variables $\Upsilon$ for \emph{branching single node component} of network of Fig.~\ref{fig:fig6}.}\label{fig:UpsilonComputationOfFig6}
\end{figure}

In fact, this problem is originated from the possibility of attaining both states of 0 and 1 in the same time. Thus, to deal with this problem in our proposed algorithm, we consider sequences of states that this node attain the desired state. For each of these sequence of states, we consider possibility for other nodes to attain desired states. For example in network of Fig.~\ref{fig:fig6}, once we assign $\Upsilon^{v_1}_0(1)=true$ and $\Upsilon^{v_1}_1(1)=false$, then we check remaining nodes $v_2$ and $v_3$ whether they are able to attain desirable states or not. If not, we return and set $\Upsilon^{v_1}_0(1)=false$ and $\Upsilon^{v_1}_1(1)=true$. If for all the sequence states of branching nodes, the remaining nodes are not able to attain the desirable state, we announce that there is no proper control sequence.

\subsubsection{Multi-Node Components}
In this case, current component consists of at least two nodes.
We apply Datta algorithm in this case~\cite{datta2003external}. Datta et al form a table $D[v_1(t), v_2(t), \dots, v_n(t), t]$, for $t=\tau$ to $t=0$ according to the following procedure:

\begin{eqnarray}
D[v_1(t), v_2(t), \dots, v_n(t), \tau] &=& \begin{cases}
    1,& \text{\parbox{8cm}{\raggedright if $[v_1(t), v_2(t), \dots, v_n(t)]=v^{\tau}$}}\\
    0,& \text{otherwise}
  \end{cases}\\
D[v_1(t-1), v_2(t-1), \dots, v_n(t-1), t-1] &=& \begin{cases}
    1,& \text{\parbox{8cm}{\raggedright if there exists $(v^t,u)$ such that $D[v_1(t), v_2(t), \dots, v_n(t), t]=1$ and $v^t=f(v^{t-1},u)$}}\\
    0,& \text{otherwise}
  \end{cases}
\end{eqnarray}
Then, there exists a desired control sequence if and only if $D[v_1(t), v_2(t), \dots, v_n(t), 0]=1$ holds for $v^0$.

After applying Datta algorithm, we check if it is possible to attain the desirable state in time step $t=\tau$. Then,
if it is possible, for a node to attain a state, we fill $\Upsilon$ array accordingly.
For example, node $v_i$ in time step $t$ may attain desirable state in time step $\tau$ with both states 0 and 1, thus, we would have $\Upsilon^{v_i}_0(t)=true$ and $\Upsilon^{v_i}_1(t)=true$. Then we partition nodes with edges going out of the component into two categories, nodes with at most one outgoing edge and nodes with at least two outgoing edges (branching node). Nodes with at most one outgoing edge are easy to handle, however, we treat branching nodes which may attain both states 0 and 1 in the same time step, like \emph{branching single node components}.

\subsubsection{Steps of Proposed Algorithm}
According to the above-mentioned issues and problems, the proposed algorithm steps are as follows:
\begin{enumerate}
  \item Divide the network graph into strongly connected components. The resulting graph would be a DAG\footnote{Directed Acyclic Graph}.
  \item Meet the components based on a topological order.
    \begin{enumerate}
      \item If related component is a \emph{non-branching single node component}:
      \begin{enumerate}
        \item Taking benefit from the dynamic programming, form $\Upsilon$ array of the related node till the desirable time step $\tau$. (Since the nodes are met based on a topological order, and all the states of the entering nodes of the related node till the time step $\tau$ are calculated before, this would be possible).
        \item In time step $\tau$, check if the related node has attained to the desirable state or not; if the answer was negative, go to step 4.
      \end{enumerate}
      \item If the related component is a \emph{branching single node component}:
      \begin{enumerate}
        \item Taking benefit from the dynamic programming, form $\Upsilon$ array of the related node till the desirable time step $\tau$.
        \item Check if this node in the same time steps has the possibility of attaining both the states 0 and 1 or not:
        \begin{enumerate}
          \item If the answer was negative treat this node like the nodes of \emph{non-branching single node component}.
          \item If the answer was positive, first check if it is possible to attain the desirable state in time step $\tau$, or not.
            \item If the answer was negative, go to step 4.
            \item If there existed such possibility, consider a sequence of states that this node can attain the desirable state.
        \end{enumerate}
      \end{enumerate}
      \item If related component is a \emph{multi-node component}:
      \begin{enumerate}
        \item Form the $D[v_1(t), v_2(t), \dots, v_n(t), t]$ table, for $t=\tau$ to $t=0$.
        \item Check if it is possible to attain the desirable state in time step $t=\tau$:
        \begin{enumerate}
          \item If the answer was negative, go to step 4.
          \item If it would be possible to attain the desirable state in time step $t=\tau$, for each node based on sequence of states that there is the possibility to attain a desirable state, form the $\Upsilon$ array of that node. For example, of the node $v_i$ in time step $t$ could attain the desirable state in time step $t=\tau$ with both states of 0 and 1, then we would have $\Upsilon^{v_i}_0(t)=true$ and $\Upsilon^{v_i}_1(t)=true$.
          \item From among the nodes which are edged out of the component, if there is a branching node with the possibility to attain both the states of 0 and 1 in the same time steps, consider a sequence of states that this branching node can attain the desirable state.
        \end{enumerate}
      \end{enumerate}
    \end{enumerate}
    \item Announce that there is the control sequence and terminate the algorithm.
    \item Return on topological order and consider another of sequence state for branching node; if there is not branching node or if consider all sequence state of previous branching nodes, announce that there is no control sequence and terminate the algorithm.
\end{enumerate}

\begin{figure}
  \includegraphics[width=6in]{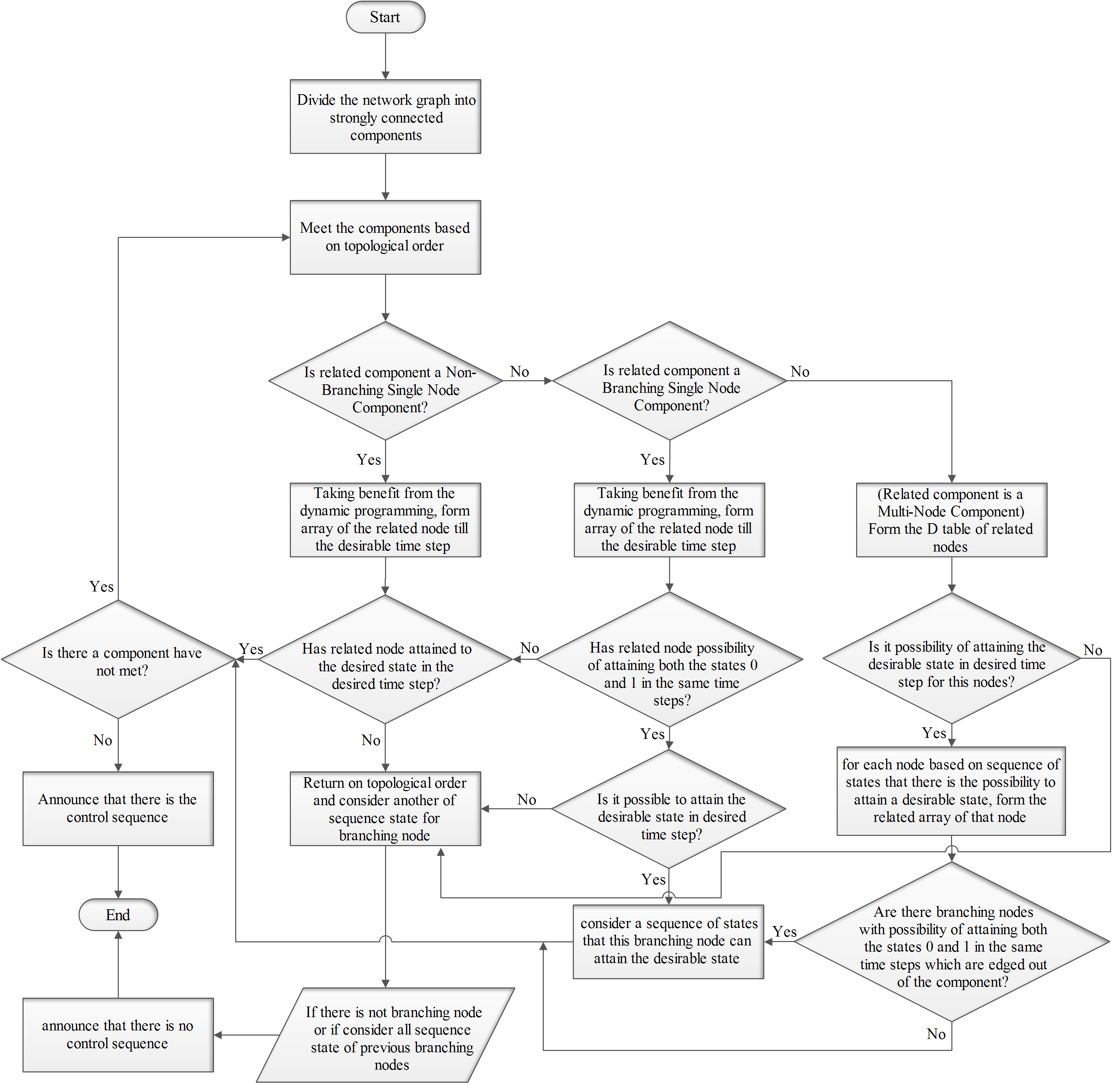}
  \centering
  \caption{flowchart of proposed algorithm.}\label{fig:fig9}
\end{figure}

The flowchart of proposed algorithm as it is shown in Fig.~\ref{fig:fig9}

\section{Application to Biomolecular Networks}
\begin{figure}
  \includegraphics[width=5in]{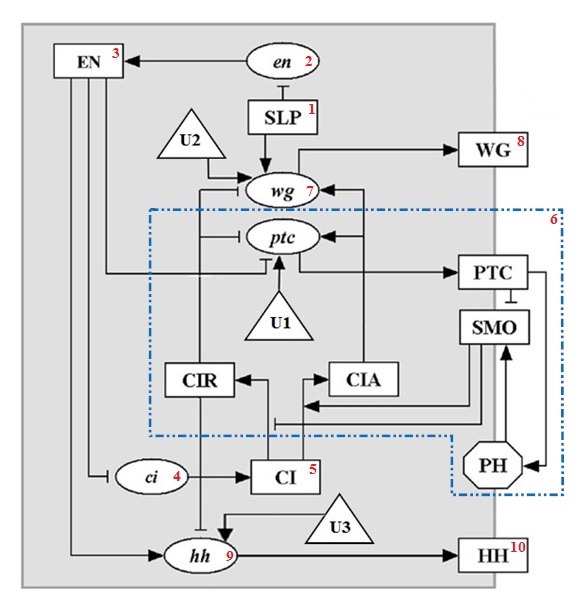}
  \centering
  \caption{Boolean network model of Drosophila melanogaster. Dotted lines indicate a \emph{multi-node component}, and the numberings beside nodes indicate topological orders.}\label{fig:fig7}
\end{figure}

In this section, the steps of applying the proposed algorithm on Drosophila melanogaster network~\cite{albert2004boolean} are reviewed. for a cell including 15 genes. As it can be seen in Fig.~\ref{fig:fig7}, three external nodes of U1, U2 and U3 are added to this network (these three nodes are marked as triangles). Evolving functions regarding the added control sequences are also mentioned in Tab.~\ref{tab:tab1}.

\def\AND{\wedge}
\def\OR{\vee}
\def\NOT{\neg}
\begin{table}
  \caption{Evolution functions for the Boolean network model of Drosophila melanogaster.}\label{tab:tab1}
  \[
  \begin{array}{ll} \hline
   \text{Node} & \text{Boolean updating function} \\ \hline \hline
   \text{SLP} & \text{SLP}^{t+1} = \text{SLP}^t \\
   \text{wg} & \text{wg}^{t+1}= ((\text{CIA}^t \AND \text{SLP}^t \AND \NOT \text{CIR}^t) \OR (\text{wg}^t \AND (\text{CIA}^t \OR \text{SLP}^t) \AND \NOT \text{CIR}^t)) \AND \text{U2}^t \\
  \text{WG} & \text{WG}^{t+1}= \text{wg}^t \\
  \text{en} & \text{en}^{t+1} = \NOT \text{SLP}^t \\
  \text{EN} & \text{EN}^{t+1} = \text{en}^t \\
  \text{hh} & \text{hh}^{t+1} = \text{EN}^t \AND \NOT \text{CIR}^t \AND \text{U3}^t \\
  \text{HH} & \text{HH}^{t+1} = \text{hh}^t \\
  \text{ptc} & \text{ptc}^{t+1} = \text{CIA}^t \AND \NOT \text{EN}^t \AND \NOT \text{CIR}^t \AND \text{U1}^t \\
  \text{PTC} & \text{PTC}^{t+1} = \text{ptc}^t \OR \text{PTC}^t \\
  \text{PH} & \text{PH}^{t+1} = \text{PTC}^t \\
  \text{SMO} & \text{SMO}^{t+1} = \NOT \text{PTC}^t \\
  \text{ci} &\text{ci}ci^{t+1} = \NOT \text{EN}^t \\
  \text{CI} &  \text{CI}^{t+1} = \text{ci}^t \\
  \text{CIA} & \text{CIA}^{t+1} =  \text{CI}^t \AND \text{SMO}^t \\
  \text{CIR} & \text{CIR}^{t+1} =  \text{CI}^t \AND \NOT \text{SMO}^t \\ \hline
  \end{array}
  \]
\end{table}

First, we divided network's graph into strongly connected components. As it could be seen in Fig.~\ref{fig:fig7}, graph has only one \emph{multi-node component}, which is shown through dotted lines.
Then, nodes are ordered based on topological sort and this sort is depicted in Fig.~\ref{fig:fig7} with numberings beside each node.

 we treat graph nodes one by one according to their topological orderings. First, ``SLP'' node is met. This node is a \emph{none-branching single node component}; therefore, we compute $\Upsilon$ for this node till $t = \tau$ . Since this node is a constant node (a node without entering edge), it will always keep its initial state. The $\Upsilon$ concerning this node will be filled to this constant value. Then, for time step $\tau$, we check that if this node has attained the desirable state or not. If not, it would be announced that there existed no control sequence, and the algorithm terminates.

Then the second node in topological sort that is the ``en'' node will be visited. This node is also a \emph{none-branching single node component}, thus, it would be acted same as for ``SLP'' node.

Then the ``EN'' node would be considered. It is a branching node, but since it has no control node before itself, it has possibility to attain only one state in every time step. Therefore this node would be treated like ``SLP'' and ``EN'' nodes. For ``ci'' and ``CI'' nodes, the very procedure will go.

Then we reach a \emph{multi-node component}. For this strongly connected component, the Datta et al algorithm is applied, then it would be checked if there is the possibility to attain the desirable state in time step $\tau$ for the nodes inside this component or not. If we can attain the desirable state within the time step $\tau$, the $\Upsilon$ for each node would be filled. Then, branching nodes which have outgoing edges from the component, would be handled as ``CIR'' node to check if theses nodes are possible to attain two states of 0 and 1 in a time step. If they could, a state sequences that we can attain the desirable state, considered and algorithm would be recalled for network remaining nodes.

Then nodes ``wg'', ``WG'', ``hh'' and ``HH'' are considered. these nodes are \emph{none-branching single node components}. Please note that, if for example ``CIR'' with two state sequences can attain the desirable state, and for the first state sequence, it would not be possible for the next nodes (e.g. ``wg'') to attain the desirable state, the algorithm will return and for the sequences of the second state of ``CIR'', it would check the remaining nodes. In the case that all the nodes were possible to attain the desirable state, it will be announced that there exist a control sequence that cause attaining the desirable state in time step $\tau$.

\begin{figure}
  \includegraphics[width=5in]{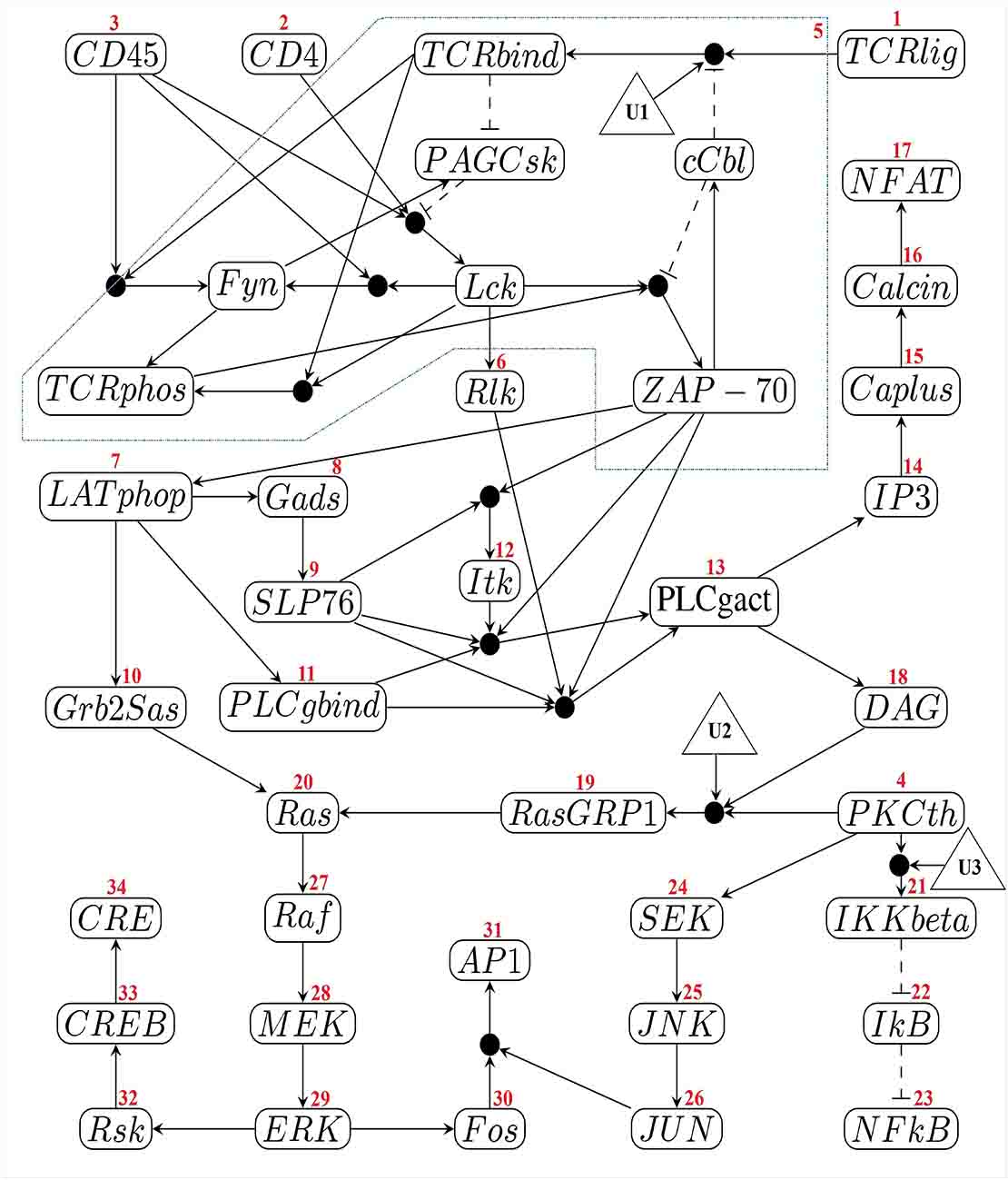}
  \centering
  \caption{Boolean network model of T-cell receptor. Network's graph has one \emph{multi-node component}. The numberings beside each node indicate their topological ordering.}\label{fig:fig8}
\end{figure}

And also, the steps of the proposed algorithm were applied on Boolean network model of T-cell receptor kinetics~\cite{klamt2006methodology}, and the \emph{multi-node component} together with the order of nodes based on a topological sort are depicted in Fig.~\ref{fig:fig8}. As it is obvious in this figure, this model of Boolean network has 40 genes and one \emph{multi-node component}. Three external nodes U1, U2 and U3, are added to the Boolean network.

In T-cell receptor kinetics, network which can be seen in Fig.~\ref{fig:fig8}, arrows with pointed heads represent activation and dashed arrows with bar heads represent inhibition (the dashed arrows represent “not” which is related to that node).
The network is represented as a network of ``or''s of ``and''s. Thus, large filled circles representing “and” of their inputs, while when edges are going to a node, state of the node is determined according to the ``or'' of its incoming edges.
As an example see
\[ \text{Fyn}(t + 1) = (\text{CD45}(t) \AND \text{TCRbind}(t)) \OR (\text{CD45}(t) \AND \text{Lck}(t)) \]

Our suggested algorithm on the Boolean network model of Drosophila melanogaster, was implemented for an initial state (in time step $t = 0$) and a desired state (in time step $t = 6$), which is depicted in Tab.~\ref{tab:tab2}. It is noteworthy to say that there exists a control sequence to reach the desirable state in time step $t = 6$. Also the Datta et al algorithm was implemented in this dataset, for comparison. Note that, since this dataset has 22 edges and 15 internal nodes, the algorithm presented by Akutsu et al requires a tree which has 8 less edges from this network ($H = 8$); therefore the time complexity of the algorithm of transforming into a rooted tree regarding $t = \tau = 6$, with time complexity $O(2^{H(\tau+1)}(n+m)\tau)$ that $n$ is the number of internal nodes ($n = 15$) and $m$ is the number of external nodes ($m = 3$), is worse than the algorithm of Datta et al with time complexity of
$O(2^{(2n+m)}\tau)$.

The results of implementation on a PC with Dual-Core 2.5GHz CPU, 2G RAM are depicted in Tab.~\ref{tab:tab3}.

\begin{table}
  \caption{Initial state and desirable state of each node for Drosophila melanogaster network's graph}\label{tab:tab2}
  \centering
  \begin{tabular}{ccc}
      \hline
      Node & Initial State $(t=0)$ & Desired State $(t=6)$ \\ \hline \hline
      en & 0 & 1 \\
      EN & 0 & 1 \\
      SLP & 0 & 0 \\
      wg & 1 & 0 \\
      WG & 1 & 0 \\
      ptc & 0 & 0 \\
      PTC & 0 & 0 \\
      CIA & 0 & 0 \\
      CIR & 1 & 0 \\
      ci & 0 & 0 \\
      CI & 1 & 0 \\
      SMO & 0 & 1 \\
      PH & 0 & 0 \\
      hh & 1 & 0 \\
      HH & 0 & 1 \\ \hline
  \end{tabular}
\end{table}

\begin{table}
  \caption{Comparison of the algorithm for solving the control problem and runtimes of each one on the Drosophila melanogaster network's.}\label{tab:tab3}
  \centering
  \begin{tabular}{p{60mm}p{60mm}}
      \hline
      Algorithm & Runtime \\ \hline \hline
      Algorithm of Datta et al & 16.5 hours \\
      Algorithm of transforming the graph into a rooted tree structure & Over 2 days \\
      Proposed (our) algorithm & 0.92 Seconds \\ \hline
  \end{tabular}
\end{table}

Also our suggested algorithm and the Algorithm of Datta et al on Boolean network model of T-cell receptor kinetics~\cite{klamt2006methodology} which are shown in Fig.~\ref{fig:fig8} were tested regarding initial state (in time step $t = 0$) and desirable state (in time step $t = 5$) which are depicted in Tab.~\ref{tab:tab4} (it is noteworthy to say that there is a control sequence for the abovementioned desirable state in time step $t = 5$). 

This dataset has 40 internal nodes ($n = 40$) and 3 external nodes ($m = 3$). The algorithm presented by Akutsu et al requires a tree which has 19 less edges from this network ($H = 19$); therefore the time complexity of the algorithm of transforming into a rooted tree regarding $t = \tau = 5$, with time complexity $O(2^{H(\tau+1)}(n+m)\tau)$, is worse than the algorithm of Datta et al with time complexity of $O(2^{(2n+m)}\tau)$.
Comparison results are depicted in Tab.~\ref{tab:tab5}.

\begin{table}
  \caption{Initial states and desirable states of each node in T-cell receptor kinetics network graph}\label{tab:tab4}
  \centering
  \begin{tabular}{p{20mm}p{20mm}p{20mm}|p{20mm}p{20mm}p{20mm}} \hline
  Node & Initial State $(t=0)$ & Desired State $(t=5)$ &   Node & Initial State $(t=0)$ & Desired State $(t=5)$ \\ \hline \hline
  CD45 & 1 & 1 & Grb2Sas & 0 & 0 \\
  CD4 & 0 & 0 & PLCgbind & 0 & 0 \\
  TCRbind & 0 & 1 & DAG & 0 & 0 \\
  TCRlig & 1 & 1 & Ras & 0 & 0 \\
  PAGCsk & 0 & 1 & RasGRP1 & 0 & 0 \\
  cCbl & 0 & 0 & PKCth & 1 & 1 \\
  NFAT & 0 & 0 & CRE & 0 & 0 \\
  Fyn & 0 & 1 & Raf & 0 & 0 \\
  Lck & 1 & 0 & SEK & 0 & 1 \\
  Calcin & 0 & 0 & IKKbeta & 0 & 1 \\
  TCRphos & 0 & 1 & CREB & 1 & 0 \\
  Rlk & 0 & 0 & MEK & 0 & 1 \\
  ZAP-70 & 0 & 0 & AP1 & 0 & 0 \\
  Caplus & 0 & 0 & JNK & 0 & 1 \\
  LATphop & 0 & 0 & IKB & 1 & 0 \\
  Gads & 0 & 0 & Rsk & 0 & 0 \\
  IP3 & 0 & 0 & ERK & 0 & 0 \\
  SLP76 & 0 & 0 & Fos & 1 & 0 \\
  Itk & 0 & 0 & JUN & 0 & 1 \\
  PLCgact & 1 & 0 & NFkB & 0 & 1 \\ \hline
  \end{tabular}
\end{table}

\begin{table}
  \caption{Comparison of algorithms for solving the control problem and the execution time of each of them on the T-cell network's receptor kinetics.}\label{tab:tab5}
  \centering
  \begin{tabular}{p{60mm}p{60mm}}
      \hline
      Algorithm & Runtime \\ \hline \hline
      Algorithm of Datta et al & Over 12 days \\
      Algorithm of transforming the graph into a rooted tree structure & Over 27 days \\
      Proposed (our) algorithm & 1.83 Seconds \\ \hline
  \end{tabular}
\end{table}

\section{Conclusions}
In this research we tried to improve running time of solutions of the gene regulatory network control problem. The extent of improvements and efficiency of our proposed algorithm depends on the size of the \emph{multi-node components} of the network and also on the positioning of the control nodes or more generally, on the accessibility of both states of 0 and 1 in the same time steps for branching nodes.

Since in the proposed algorithm, the nodes are met based on a topological order, the states of the entering nodes of each node are calculated before visiting that node. As a result, in each node it would be possible to handle states from initial time step to desirable time. This would cause an earlier detection if a node is not possible to attain the desirable state (there be no need to check all nodes to time step $\tau$), and as a result, the lack of a control sequence is reported earlier. Despite the high improvement of the suggested algorithm in solving the control problem, if all the nodes of the network's graph are among the strongly connected components, the suggested algorithm turns into the algorithm of Datta et al. This might happen very rarely. Meanwhile, handling this problem can be considered as the future work.

\bibliographystyle{plain}
\bibliography{BNControl-Moradi}

\begin{thebibliography}{10}

\bibitem{akutsu2007control}
Tatsuya Akutsu, Morihiro Hayashida, Wai-Ki Ching, and Michael~K Ng.
\newblock Control of boolean networks: hardness results and algorithms for tree
  structured networks.
\newblock {\em Journal of Theoretical Biology}, 244(4):670--679, 2007.

\bibitem{akutsu2000inferring}
Tatsuya Akutsu, Satoru Miyano, and Satoru Kuhara.
\newblock Inferring qualitative relations in genetic networks and metabolic
  pathways.
\newblock {\em Bioinformatics}, 16(8):727--734, 2000.

\bibitem{albert2004boolean}
R{\'e}ka Albert.
\newblock Boolean modelingof genetic regulatory networks.
\newblock In {\em Complex networks}, pages 459--481. Springer, 2004.

\bibitem{albert2000dynamics}
R{\'e}ka Albert and Albert-L{\'a}szl{\'o} Barab{\'a}si.
\newblock Dynamics of complex systems: Scaling laws for the period of boolean
  networks.
\newblock {\em Physical Review Letters}, 84(24):5660, 2000.

\bibitem{alberts2002molecular}
Bruce Alberts, A~Johnson, J~Lewis, M~Raff, K~Roberts, and P~Walter.
\newblock Molecular biology of the cell. 2008, new york: Garland science.
\newblock {\em Google Scholar}, 2002.

\bibitem{amaral2004emergence}
Lu{\'\i}s~AN Amaral, Albert D{\'\i}az-Guilera, Andre~A Moreira, Ary~L
  Goldberger, and Lewis~A Lipsitz.
\newblock Emergence of complex dynamics in a simple model of signaling
  networks.
\newblock {\em Proceedings of the National Academy of Sciences of the United
  States of America}, 101(44):15551--15555, 2004.

\bibitem{bar2004analyzing}
Ziv Bar-Joseph.
\newblock Analyzing time series gene expression data.
\newblock {\em Bioinformatics}, 20(16):2493--2503, 2004.

\bibitem{batchelor2009ups}
Eric Batchelor, Alexander Loewer, and Galit Lahav.
\newblock The ups and downs of p53: understanding protein dynamics in single
  cells.
\newblock {\em Nature Reviews Cancer}, 9(5):371--377, 2009.

\bibitem{bornholdt2008boolean}
Stefan Bornholdt.
\newblock Boolean network models of cellular regulation: prospects and
  limitations.
\newblock {\em Journal of the Royal Society Interface}, 5(Suppl 1):S85--S94,
  2008.

\bibitem{choi2012attractor}
Minsoo Choi, Jue Shi, Sung~Hoon Jung, Xi~Chen, and Kwang-Hyun Cho.
\newblock Attractor landscape analysis reveals feedback loops in the p53
  network that control the cellular response to dna damage.
\newblock {\em Sci. Signal.}, 5(251):ra83--ra83, 2012.

\bibitem{cormen2009introductiontoalgorithm}
Thomas~H Cormen.
\newblock {\em Introduction to algorithms}.
\newblock MIT press, 2009.

\bibitem{datta2003external}
Aniruddha Datta, Ashish Choudhary, Michael~L Bittner, and Edward~R Dougherty.
\newblock External control in markovian genetic regulatory networks.
\newblock {\em Machine learning}, 52(1-2):169--191, 2003.

\bibitem{datta2004external}
Aniruddha Datta, Ashish Choudhary, Michael~L Bittner, and Edward~R Dougherty.
\newblock External control in markovian genetic regulatory networks: the
  imperfect information case.
\newblock {\em Bioinformatics}, 20(6):924--930, 2004.

\bibitem{erler2012network}
Janine~T Erler and Rune Linding.
\newblock Network medicine strikes a blow against breast cancer.
\newblock {\em Cell}, 149(4):731--733, 2012.

\bibitem{gao2013principle}
Bo~Gao, Lixiang Li, Haipeng Peng, J{\"u}rgen Kurths, Wenguang Zhang, and Yixian
  Yang.
\newblock Principle for performing attractor transits with single control in
  boolean networks.
\newblock {\em Physical Review E}, 88(6):062706, 2013.

\bibitem{geva2006oscillations}
Naama Geva-Zatorsky, Nitzan Rosenfeld, Shalev Itzkovitz, Ron Milo, Alex Sigal,
  Erez Dekel, Talia Yarnitzky, Yuvalal Liron, Paz Polak, Galit Lahav, et~al.
\newblock Oscillations and variability in the p53 system.
\newblock {\em Molecular systems biology}, 2(1), 2006.

\bibitem{harris2002model}
Stephen~E Harris, Bruce~K Sawhill, Andrew Wuensche, and Stuart Kauffman.
\newblock A model of transcriptional regulatory networks based on biases in the
  observed regulation rules.
\newblock {\em Complexity}, 7(4):23--40, 2002.

\bibitem{helikar2011boolean}
Tom{\'a}{\v{s}} Helikar, Naomi Kochi, John Konvalina, and Jim~A Rogers.
\newblock Boolean modeling of biochemical networks.
\newblock {\em The Open Bioinformatics Journal}, 5:16--25, 2011.

\bibitem{helikar2008emergent}
Tom{\'a}{\v{s}} Helikar, John Konvalina, Jack Heidel, and Jim~A Rogers.
\newblock Emergent decision-making in biological signal transduction networks.
\newblock {\em Proceedings of the National Academy of Sciences},
  105(6):1913--1918, 2008.

\bibitem{joh2011lyse}
Richard~I Joh and Joshua~S Weitz.
\newblock To lyse or not to lyse: transient-mediated stochastic fate
  determination in cells infected by bacteriophages.
\newblock {\em PLoS Comput Biol}, 7(3):e1002006, 2011.

\bibitem{stuart1993origins}
Stuart~A. Kauffman.
\newblock {\em The origins of order: Self organization and selection in
  evolution}.
\newblock Oxford University Press, USA, 1993.

\bibitem{kim2011reduction}
Jeong-Rae Kim, Junil Kim, Yung-Keun Kwon, Hwang-Yeol Lee, Pat Heslop-Harrison,
  and Kwang-Hyun Cho.
\newblock Reduction of complex signaling networks to a representative kernel.
\newblock {\em Science signaling}, 4(175):ra35--ra35, 2011.

\bibitem{kim2013discovery}
Junil Kim, Sang-Min Park, and Kwang-Hyun Cho.
\newblock Discovery of a kernel for controlling biomolecular regulatory
  networks.
\newblock {\em Scientific reports}, 3, 2013.

\bibitem{kitano2002computational}
Hiroaki Kitano.
\newblock Computational systems biology.
\newblock {\em Nature}, 420(6912):206--210, 2002.

\bibitem{kitano2004cancer}
Hiroaki Kitano.
\newblock Cancer as a robust system: implications for anticancer therapy.
\newblock {\em Nature Reviews Cancer}, 4(3):227--235, 2004.

\bibitem{klamt2006methodology}
Steffen Klamt, Julio Saez-Rodriguez, Jonathan~A Lindquist, Luca Simeoni, and
  Ernst~D Gilles.
\newblock A methodology for the structural and functional analysis of signaling
  and regulatory networks.
\newblock {\em BMC bioinformatics}, 7(1):56, 2006.

\bibitem{langmead2009symbolic}
Christopher~James Langmead and Sumit~Kumar Jha.
\newblock Symbolic approaches for finding control strategies in boolean
  networks.
\newblock {\em Journal of Bioinformatics and Computational Biology},
  7(02):323--338, 2009.

\bibitem{lee2012sequential}
Michael~J Lee, S~Ye Albert, Alexandra~K Gardino, Anne~Margriet Heijink, Peter~K
  Sorger, Gavin MacBeath, and Michael~B Yaffe.
\newblock Sequential application of anticancer drugs enhances cell death by
  rewiring apoptotic signaling networks.
\newblock {\em Cell}, 149(4):780--794, 2012.

\bibitem{li2004yeast}
Fangting Li, Tao Long, Ying Lu, Qi~Ouyang, and Chao Tang.
\newblock The yeast cell-cycle network is robustly designed.
\newblock {\em Proceedings of the National Academy of Sciences of the United
  States of America}, 101(14):4781--4786, 2004.

\bibitem{liang1998reveal}
Shoudan Liang, Stefanie Fuhrman, and Roland Somogyi.
\newblock Reveal, a general reverse engineering algorithm for inference of
  genetic network architectures.
\newblock 1998.

\bibitem{liu2014controllability}
Yang Liu, Hongwei Chen, and Bo~Wu.
\newblock Controllability of boolean control networks with impulsive effects
  and forbidden states.
\newblock {\em Mathematical Methods in the Applied Sciences}, 37(1):1--9, 2014.

\bibitem{murrugarra2012modeling}
David Murrugarra, Alan Veliz-Cuba, Boris Aguilar, Seda Arat, and Reinhard~C
  Laubenbacher.
\newblock Modeling stochasticity and variability in gene regulatory networks.
\newblock {\em EURASIP J. Bioinformatics and Systems Biology}, 2012:5, 2012.

\bibitem{pal2005intervention}
Ranadip Pal, Aniruddha Datta, Michael~L Bittner, and Edward~R Dougherty.
\newblock Intervention in context-sensitive probabilistic boolean networks.
\newblock {\em Bioinformatics}, 21(7):1211--1218, 2005.

\bibitem{pandey2010boolean}
Sona Pandey, Rui-Sheng Wang, Liza Wilson, Song Li, Zhixin Zhao, Timothy~E
  Gookin, Sarah~M Assmann, and R{\'e}ka Albert.
\newblock Boolean modeling of transcriptome data reveals novel modes of
  heterotrimeric g-protein action.
\newblock {\em Molecular systems biology}, 6(1), 2010.

\bibitem{poret2014silico}
Arnaud Poret and Jean-Pierre Boissel.
\newblock An in silico target identification using boolean network attractors:
  avoiding pathological phenotypes.
\newblock {\em Comptes rendus biologies}, 337(12):661--678, 2014.

\bibitem{qiu2014control}
Yushan Qiu, Takeyuki Tamura, Wai-Ki Ching, and Tatsuya Akutsu.
\newblock On control of singleton attractors in multiple boolean networks:
  integer programming-based method.
\newblock {\em BMC systems biology}, 8(1):1, 2014.

\bibitem{saadatpour2011dynamical}
Assieh Saadatpour, Rui-Sheng Wang, Aijun Liao, Xin Liu, Thomas~P Loughran,
  Istv{\'a}n Albert, and R{\'e}ka Albert.
\newblock Dynamical and structural analysis of a t cell survival network
  identifies novel candidate therapeutic targets for large granular lymphocyte
  leukemia.
\newblock {\em PLoS Comput Biol}, 7(11):e1002267, 2011.

\bibitem{shih2004prediction}
KC~Shih, RM~Chen, RM~Hu, FM~Liu, HK~Chen, and Jeffrey~JP Tsai.
\newblock Prediction of gene regulatory networks using differential expression
  of cdna microarray data.
\newblock In {\em Multimedia Software Engineering, 2004. Proceedings. IEEE
  Sixth International Symposium on}, pages 378--385. IEEE, 2004.

\bibitem{tyson2001network}
John~J Tyson, Kathy Chen, and Bela Novak.
\newblock Network dynamics and cell physiology.
\newblock {\em Nature Reviews Molecular Cell Biology}, 2(12):908--916, 2001.

\bibitem{vera2013ocsana}
Paola Vera-Licona, Eric Bonnet, Emmanuel Barillot, and Andrei Zinovyev.
\newblock Ocsana: optimal combinations of interventions from network analysis.
\newblock {\em Bioinformatics}, 29(12):1571--1573, 2013.

\bibitem{wang2013therapeutic}
Wei Wang.
\newblock Therapeutic hints from analyzing the attractor landscape of the p53
  regulatory circuit.
\newblock {\em Science signaling}, 2013.

\bibitem{yousefi2013optimal}
Mohammadmahdi~R Yousefi, Amitava Datta, and Edward Dougherty.
\newblock Optimal intervention in markovian gene regulatory networks with
  random-length therapeutic response to antitumor drug.
\newblock {\em Biomedical Engineering, IEEE Transactions on},
  60(12):3542--3552, 2013.

\bibitem{yousefi2012optimal}
Mohammadmahdi~R Yousefi, Aniruddha Datta, and Edward~R Dougherty.
\newblock Optimal intervention strategies for therapeutic methods with
  fixed-length duration of drug effectiveness.
\newblock {\em IEEE Transactions on Signal Processing}, 60(9):4930--4944, 2012.

\bibitem{yousefi2013intervention}
Mohammadmahdi~R Yousefi and Edward~R Dougherty.
\newblock Intervention in gene regulatory networks with maximal phenotype
  alteration.
\newblock {\em Bioinformatics}, 29(14):1758--1767, 2013.

\bibitem{yousefi2014comparison}
Mohammadmahdi~R Yousefi and Edward~R Dougherty.
\newblock A comparison study of optimal and suboptimal intervention policies
  for gene regulatory networks in the presence of uncertainty.
\newblock {\em EURASIP J. Bioinformatics and Systems Biology}, 2014:6, 2014.

\bibitem{zeng2010decision}
Lanying Zeng, Samuel~O Skinner, Chenghang Zong, Jean Sippy, Michael Feiss, and
  Ido Golding.
\newblock Decision making at a subcellular level determines the outcome of
  bacteriophage infection.
\newblock {\em Cell}, 141(4):682--691, 2010.

\bibitem{zhang2008network}
Ranran Zhang, Mithun~Vinod Shah, Jun Yang, Susan~B Nyland, Xin Liu, Jong~K Yun,
  R{\'e}ka Albert, and Thomas~P Loughran.
\newblock Network model of survival signaling in large granular lymphocyte
  leukemia.
\newblock {\em Proceedings of the National Academy of Sciences},
  105(42):16308--16313, 2008.

\end{thebibliography}

\end{document}